\documentclass[twocolumn,showpacs,nofootinbib,10pt]{revtex4-1}

\usepackage{color}
\usepackage{graphicx}
\usepackage{graphicx,float}\usepackage[all]{xy}
\usepackage{amsmath,upgreek,tikz,bm}
 \newcommand{\blt}{\textcolor{black}}
\usepackage{amssymb}

\usepackage{textgreek}

\newcommand{\beq}{\begin{eqnarray}}
\newcommand{\eeq}{\end{eqnarray}}
\newcommand{\be}{\begin{equation}}
\newcommand{\ee}{\end{equation}}

\newcommand{\bea}{\begin{eqnarray}}
\newcommand{\eea}{\end{eqnarray}}

\newcommand{\ba}{\begin{eqnarray}}
\newcommand{\ea}{\end{eqnarray}}
\newcommand\orcidroldao{{\href{https://orcid.org/0000-0003-3978-532X}{\orcidicon}}}
\newcommand{\orcidicon}{%
	\begin{tikzpicture}
	\draw[lime, fill=lime] (0,0)
		circle [radius=0.16]
		node[white] {{\fontfamily{qag}\selectfont \tiny ID}};
	\draw[white, fill=white] (-0.0625,0.095)
		circle [radius=0.007];
	\end{tikzpicture}	\hspace{-2mm}
}

\usepackage[colorlinks,hyperindex,unicode]{hyperref}
\definecolor{green1}{RGB}{0,128,0} 
\hypersetup{hidelinks,backref=true,pagebackref=true,hyperindex=true,colorlinks=true,breaklinks=true,urlcolor= blue}
\hypersetup{%
  colorlinks = true,
  linkcolor  = blue,
  citecolor = cyan,
}
\usepackage{bookmark,textgreek}
\usepackage{hyperref,color,xcolor}
\hypersetup{hidelinks,hyperindex=true,colorlinks=true,breaklinks=true,urlcolor= blue}
\hypersetup{%
  colorlinks = true,
  linkcolor  = blue
}
\begin{document}
\title{Probing the minimal geometric deformation with trace and Weyl anomalies}

\author{P. Meert}
\email{pedro.meert@ufabc.edu.br} 
\affiliation{Center of Physics, Universidade Federal do ABC,  09210-580, Santo Andr\'e, Brazil.}
\author{R. da Rocha\orcidroldao\!\!}
\email{roldao.rocha@ufabc.edu.br}
\affiliation{Center of Mathematics,  Federal University of ABC, 09210-580, Santo Andr\'e, Brazil.}

%
%
%\pacs{11.25.Tq, 11.25.-w, 04.50.Gh} 

%\date{December 25, 2016}

\begin{abstract} 
The method of minimal geometric deformation (MGD) is used to derive static, strongly gravitating,  spherically symmetric, compact stellar distributions. The trace and Weyl  anomalies are then employed to probe the MGD in the holographic setup, as a realistic model, playing a prominent role in AdS/CFT. 

\end{abstract}

%\pacs{04.50.Gh, 04.70.Bw, 11.25.-w}


\maketitle

\section{Introduction} 

The minimal geometric deformation, MGD, and the MGD-decoupling methods consist of well-succeeded procedures that can construct analytical solutions of the brane Einstein's effective field equations, in AdS/CFT and its membrane paradigm  \cite{Ovalle:2017fgl,Casadio:2012rf,darkstars,Casadio:2015gea,Ovalle:2017wqi}.
 Our universe, the codimension-1 brane with intrinsic tension, is assumed to be embedded in a bulk   \cite{Antoniadis:1998ig}. 
 The minimal anisotropic procedure onto the brane has been thrivingly employed to engender exact inner solutions to Einstein's field equations for static and nonuniform stellar configurations, containing local and nonlocal bulk terms. The MGD is a formal approach that generates holographic and realistic varieties of not only any solution in General Relativistic (GR) but extended ones, on fluid branes \cite{Ovalle:2010zc}. Weyl functions into stellar distributions can create effective physical signatures of the bulk. 
 
 Gravitational field equations, in GR, take into account the regime of a rigid brane, where the  tension is infinite. However, to match recent observational data, the  brane tension, that emulates the vacuum energy itself, must be finite. The current range of the brane tension, $\sigma$, is given by  $\sigma \gtrapprox  2.813\times10^{-6} \;{\rm GeV}$ \cite{Fernandes-Silva:2019fez}. The brane tension controls the MGD of the Schwarzschild spacetime, that is a solution of the brane Einstein equations, and also drives other deformed solutions that include tidal charge \cite{Casadio:2013uma,Ovalle:2013vna,Ovalle:2019qyi,Ovalle:2013xla,Ovalle:2018vmg}. The brane tension is proportional to the universe temperature, shown by WMAP regarding the CMB anisotropy  \cite{Abdalla:2009pg,daRocha:2012pt}. The MGD was employed to construct compact stellar distributions on the brane \cite{covalle2,Ovalle:2014uwa,Ovalle:2016pwp}. Refs.  \cite{Kanno:2002iaa,Soda:2010si} introduced the bridge between braneworld models and the holographic AdS/CFT setup \cite{Kanno:2002iaa,Soda:2010si,Skenderis:2003da}.

 The MGD and extended MGD solutions were studied in Refs. \cite{Casadio:2012pu,Casadio:2015jva,Casadio:2016aum} under different phenomenological setups. Analog MGD models of gravity, using acoustics in a moving fluid, were proposed \cite{daRocha:2017lqj}. Refs. \cite{Contreras:2018gzd,Ovalle:2007bn,Ovalle:2018ans,Sharif:2018tiz,Sharif:2019mzv,Morales:2018urp,Rincon:2019jal,Hensh:2019rtb,Ovalle:2019lbs,Gabbanelli:2019txr} have paved  a robust way to construct solutions of the brane Einstein effective field equations, using  MGD methods. Moreover,  MGD-decouping procedure has been employed to engender anisotropic solutions that describe anisotropic stellar configurations  \cite{Tello-Ortiz2020,Gabbanelli:2018bhs,Panotopoulos:2018law,Heras:2018cpz,Contreras:2018vph,Ovalle:2017khx,Tello-Ortiz:2020euy}. The observation of gravitational lensing effects to detect signatures of MGD stellar configurations was proposed in  \cite{Cavalcanti:2016mbe}, and MGD stars composed of glueball condensates have been also discussed \cite{daRocha:2017cxu,Fernandes-Silva:2018abr}. 
 %In addition,   
 %MGD black holes in the GUP context  were studied in Ref. \cite{Casadio:2017sze}, and 
 MGD Dirac stars were introduced and scrutinized in Ref. \cite{daRocha:2020rda}.

The MGD-decoupling procedure iteratively constructs, upon a given isotropic source of gravitational field, anisotropic compact sources of gravity, that are weakly coupled. One starts with a perfect fluid, then  coupling it to more elaborated  stress-energy-momentum tensors that underlie  realistic compact configurations \cite{Maurya:2019kzu,PerezGraterol:2018eut,Morales:2018nmq,Contreras:2019iwm,Contreras:2019fbk,Singh:2019ktp}.  Ref. \cite{Tello-Ortiz:2019gcl} demonstrated that for positive anisotropy, when the radial pressure is smaller than the tangential pressure, the stellar distribution exerts a  force that is  repulsive and  compensates for the  pressure. In this way, anisotropic compact stars  are more plausible stellar configurations to happen, as proposed and studied in Refs.  \cite{Jasim:2018wtd,Maurya:2019sfm,Maurya:2019xcx,Maurya:2019hds,Cedeno:2019qkf,Ivanov:2018xbu,Sharif:2018toc,Stelea:2018cgm,Ovalle:2018umz,Estrada:2018zbh,Nakas:2020crd,Dev:2000gt}.  Current observations of anisotropy in compact stellar distributions, via  gravitational waves, have contributed for MGD-decoupling to occupy a successful role, as a framework that describe high density, anisotropic, astrophysical entities, that comprise X-ray sources, X-ray pulsars and X-ray bursters as well \cite{Maurya:2019noq,Maurya:2019wsk}. The MGD-decoupling has been also employed to study strange stellar configurations, as the astrophysical object SAX J1808.4-3658 \cite{Tello-Ortiz:2019gcl}, as the brightest X-ray burst ever observed, as spotted by the Neutron Star inner Composition Explorer (NICER). In addition, anisotropic neutron compact stellar configurations were used to describe the compact astrophysical objects 4U 1820.30 and 1728.34, RX J185635.3754, PSR J0348+0432 and 0943+10, for instance, \cite{Torres:2019mee,Deb:2018ccw}. Strange quark  stellar configurations were also explored in Ref. \cite{Lopes:2019psm}.
The extension of isotropic compact solutions to anisotropic ones, through the MGD-decoupling, was also proposed in Refs. \cite{Abellan:2020jjl,Tello-Ortiz:2020svg}.

\par The so-called Weyl anomaly represents the fact that conformal invariance under Weyl rescaling %, $g_{\mu\nu}\mapsto\Omega^{2}g_{\mu\nu}$, 
displayed by gravity-interacting classical fields, is no longer present after quantization. This anomalous behavior was first pointed out by Ref. \cite{Capper:1973mv}, back in 1973, and since then has been applied to many areas of theoretical physics (for a review see \cite{Duff:1993wm}). Of particular interest for this work is its application to AdS/CFT correspondence, where this anomaly is related to black holes on braneworlds, and, in this context, it is called holographic Weyl anomaly \cite{Henningson:1998gx}.
{}{Other relevant aspects of trace anomalies were comprehensively developed in the seminal Refs. \cite{Bonora:1983ff,Bonora:2017gzz,Bonora:2014qla,Kuntz:2019omq}}.

\par Comparison of the holographic Weyl anomaly to the trace anomaly of the energy-momentum tensor from 4D field theory leads to a coefficient \cite{Casadio:2003jc} that measures the back-reaction of the brane on the bulk geometry. Therefore it is a good way to understand how accurate the AdS/CFT description of the on-brane boundary theory is, given a particular solution of the codimension-1 bulk.
 
This paper is organized as follows: Sect. \ref{MGD} is dedicated  to reviewing the MGD derivation 
as a complete method to deform the Schwarzschild solution and to describe realistic stellar distributions  on finite tension branes. In Sect. \ref{dsfb}, the trace anomalies are computed for MGD solutions, from the point of view of 4D QFT, and compared to that predicted by the AdS-CFT correspondence. Sect. \ref{4} is devoted to conclusions and important perspectives.

\section{MGD: Framework and metric}
\label{MGD}
 The MGD method provides  bulk corrections to well-known solutions of GR, including high energy and nonlocal corrections  \cite{Ovalle:2013xla,darkstars}. 
 Underlying the MGD procedure, fluid branes are endowed with an intrinsic  tension, mimicking the vacuum energy \cite{Gergely:2008jr,Abdalla:2009pg}.  

For $\mu,\nu=0,1,2,3$, the brane effective Einstein field equations are given by  
\begin{equation}
\label{5d4d}
{\rm G}_{\mu\nu}
=-\Lambda_{\scalebox{.6}{brane}}\, g_{\mu\nu}+\frac{8\pi G}{c^4}\mathfrak{T}_{\mu\nu},
\end{equation} where $g_{\mu\nu}$ is the brane metric, $G$ stands for the brane Newton coupling constant and ${\rm G}_{\mu\nu}$ is the well known Einstein tensor; $\Lambda_{\scalebox{.6}{brane}}$ is  the cosmological constant on the brane, that is set to zero by fine tuning \cite{maartens}. The stress-energy-momentum tensor,  appearing  in Eq. (\ref{5d4d}), can be decomposed as \cite{GCGR}
\beq
\mathfrak{T}_{\mu\nu}
=
T_{\mu\nu}-{}{E}_{\mu\nu}+\sigma^{-1} \kappa_5^4\Uppi_{\mu\nu}
+\mathcal{K}_{\mu\nu}+M_{\mu\nu},\label{tmunu}\eeq 
where $\kappa_5^2 = 48\pi G$.
The $T_{\mu\nu}$ term represents the brane stress-energy-momentum tensor, that encodes the brane energy (including dark energy) and brane matter (including dark matter) content. Given the bulk Weyl tensor $C_{\mu\nu \rho\sigma}$, its projection onto the brane, ${}{E}_{\mu\nu}\equiv C_{\mu\nu \rho\sigma}\mathfrak{n}^\rho\mathfrak{n}^\sigma$, where $\mathfrak{n}^\sigma$ is a unitary vector field out of the brane, brings nonlocal ingredients to the  brane effective Einstein field equations (\ref{5d4d}) and, in general, is $\sigma^{-1}$-dependent. When the brane is infinitely rigid, corresponding to the GR $\sigma\to\infty$ case, the tensor ${}{E}_{\mu\nu}$ is equal to zero. It is worth to mention that the brane Weyl tensor 
contains the so-called Weyl functions, namely, the Weyl scalar, $\mathcal{U}$, and the anisotropy, $\mathcal{P}$, encoded into any stellar
configuration that is solution of (\ref{5d4d}), being both proportional to
the stellar
configuration compactness. The Weyl functions arise from the MGD undertaken by the $g_{rr}$  component of the metric, due to AdS bulk effects. In addition, generalized models, modifying the pressure 
by bulk effects, also encompass 
nonlocal terms encoding the bulk Weyl curvature \cite{Casadio:2013uma}.  The $\Uppi_{\mu\nu}$ component of the stress-energy-momentum tensor encrypts quadratic terms involving the stress-energy tensor, arising from the extrinsic curvature terms in the Einstein tensor projection onto the brane.  In fact, the brane matching conditions, applied to the extrinsic curvature tensor, $K_{\mu\nu}$, makes it to be expressed as $K_{\mu\nu}=-\kappa_5^2(\Uppi_{\mu\nu}-\Uppi g_{\mu\nu}/3)$, where $\Uppi= \Uppi_\mu^{\;\mu}$ \cite{GCGR}. Also denoting $T= T_\mu^{\;\mu}$, one can explicitly write 
\beq
\!\!\!\!\Uppi_{\mu\nu}\!=\!\left(\frac18T^{\rho\sigma}T_{\rho\sigma}\!-\!\frac{1}{4!}T^2\right)g_{\mu\nu}\!+\!\frac1{12}TT_{\mu\nu}\!-\!\frac14 T_{\mu\rho}T_{\nu}^{\;\rho}.
\eeq
The tensor $\mathcal{K}_{\mu\nu}$ in Eq. (\ref{tmunu}) describes an eventual asymmetric embedding to the brane into the AdS bulk and the $M_{\mu\nu}$ tensor includes bulk gravitons and moduli fields \cite{Gergely:2008jr,maartens}.

\par The solution considered in this work has $\mathcal{K}_{\mu\nu}=M_{\mu\nu}=0$ in Eq. \eqref{tmunu}, and the corresponding energy-momentum tensor on the brane, $T_{\mu\nu}$, is that of a perfect fluid for the inner region of the star, $r<R$, and vanishes for $r>R$, as is discussed below.

Compact stellar configurations represent analytical solutions of the gravitational field equations (\ref{5d4d}), with metric 
\begin{equation}\label{abr}
ds^{2} = -A(r) dt^{2} + {B(r)}^{-1}dr^{2} + r^2\,d\Upomega^2,
\end{equation} 
where $d\Upomega^2$ denotes the solid angle element. 
One denotes $A(r)=e^{\bm{\upnu}(r)}$ and $B(r)=e^{\upxi(r)}$, for the sake of conciseness. 

We define the 
integral  \be
\label{I}
\mathcal{I}(r)
=
\int_0^r\frac{2\mathfrak{r}^2\bm{\upnu}''(\mathfrak{r})+({\mathfrak{r}{\bm{\upnu}'(\mathfrak{r})}}+{2})^2}
{{\mathfrak{r}^2\bm{\upnu}'(\mathfrak{r})}+{4\mathfrak{r}}}\,d\mathfrak{r}
\,,
\ee
denoting by a prime the derivative with respect to the radial coordinate.

The MGD asserts that the ${B(r)}^{-1}$ radial component  of the metric (\ref{abr}) can be deformed as  \cite{Casadio:2013uma} 
\begin{eqnarray}
\label{edlrwssg}
e^{-\upxi(r)}
&=&
\upmu(r)+\upkappa(r,\sigma^{-1})
\ ,
\end{eqnarray}
where \cite{Ovalle:2010zc}
\begin{eqnarray}\label{mmm}
\!\!\!\!\!\!\upkappa(r,\sigma)={e^{-\mathcal{I}(r)}\left[\mathfrak{b}(r,\sigma)+\mathcal{J}(r)\right]},
\end{eqnarray}
for
\beq\label{iint}
\!\!\!\!\!\!\!\!\!\!\!\mathcal{J}(r)=\!\int_0^r\!\frac{2\mathfrak{r}\,e^\mathcal{I}(\mathfrak{r})}{{\mathfrak{r}\bm{\upnu}'(\mathfrak{r})}+{4}}
\!\left[L(\mathfrak{r})+\frac{\uprho(\mathfrak{r})}{G^2\sigma}\left(\uprho(\mathfrak{r})+3p(\mathfrak{r})\right)\right]\,d\mathfrak{r}.
\eeq
In Eq. (\ref{iint}), $p(\mathfrak{r})$ denotes the pressure and $\uprho(\mathfrak{r})$ the density of the compact star. 
In the limit of an infinitely rigid brane, $\sigma\to\infty$ case, the deformation reads $\upkappa(r)\to0$, recovering General Relativity.
Moreover,   the function appearing in Eq. (\ref{edlrwssg}) reads 
\begin{eqnarray}
\upmu(r) = \begin{cases} 1-{2\,{GM_{\scalebox{.6}{Schw}}}}/{c^2r}
\ ,
&
\quad r>{\rm R}\\
1-\frac{1}{G^2\,r}\int_0^r \uprho(\mathfrak{r})\mathfrak{r}^2\, d\mathfrak{r}
\,,
&
\quad r\,\leq\,{\rm R}\,,
\end{cases}
\end{eqnarray}
where ${\rm R}$ denotes the star surface radius and $M_{\scalebox{.6}{Schw}}$ is the Schwarzschild star GR mass. 
The function $L(r)=L(\bm{\upnu}(r), p(r),\uprho(r))$ encodes bulk-induced anisotropy. 
The function $\mathfrak{b}=\mathfrak{b}(r,\sigma)$ in Eq. (\ref{mmm})  will be derived soon.  The MGD $\upkappa(r)$,  in vacuum, where $p(r)=\uprho(r)=0$, will be hereon represented by $\tilde{h}(r)$ \cite{Casadio:2012rf}:
\begin{equation}
\label{def}
\tilde{h}(r,\sigma)
=\mathfrak{b}(r,\sigma)\,e^{-\mathcal{I}(r)}
\ .
\end{equation}
The MGD can be split into  terms that are factors of $\sigma^{-1}$, encompassing high energy terms, and terms that encode nonlocal effects of the Weyl fluid.
Junction conditions match the inner MGD metric, meaning that $r<{\rm R}$, for $\tilde\upkappa(r,\sigma)$ given in Eq. (\ref{edlrwssg}), making $L(r)=0$, 
in the exterior region, $r>{\rm R}$.

The Weyl fluid, that moistens the brane, can be represented by Weyl functions. The on-brane Weyl tensor, being inversely proportional to the brane tension, can be split off as    
\begin{eqnarray}
\!\!\!\!\!\!\!\!{E}_{\mu\nu} \!=\!-2\sigma^{-1}\!\left[\mathfrak{Q}_{(\mu} \mathfrak{u}_{\nu)}\!+ \mathcal{U}\!\left(\!\mathfrak{u}_\mu \mathfrak{u}_\nu \!+\! \frac{1}{3}{\rm h}_{\mu\nu}\!\right)\!+\mathcal{P}_{\mu\nu}\right], \label{A4}
\end{eqnarray}
\noindent where the vector field  $\mathfrak{u}^\mu$ is the velocity  that describes the flow of the Weyl fluid and the tensor ${\rm h}_{\mu\nu}=g_{\mu\nu}+\mathfrak{u}_\mu\mathfrak{u}_\nu$ projects quantities orthogonally to the flow direction. In addition, the $\sigma$-dependent Weyl scalar, explicitly given  by $\mathcal{U}=-\frac12\sigma{E}_{\mu\nu} \mathfrak{u}^{\mu} \mathfrak{u}^{\nu}$ represents the energy density, whereas  $\mathcal{P}_{\mu\nu}=-\frac12\sigma\left({\rm h}_{(\mu}^{\;\uprho}{\rm h}_{\nu)}^{\;\sigma}-\frac13 {\rm h}^{\uprho\sigma}{\rm h}_{\mu\nu}\right){E}_{\rho\sigma}$ denotes the  (nonlocal) anisotropic energy-stress-momentum tensor. Besides, the brane nonlocal  flux of energy is given by $\mathfrak{Q}_\upalpha = -\frac12\sigma h^{\;\rho}_{\upalpha}{E}_{\rho\upbeta}\mathfrak{u}^\upbeta$.

The trace of the anisotropic energy-stress-momentum tensor and the Weyl scalar in the outer region are, respectively, given by
\ba
\label{pp2}
\,{{}{\mathcal{P}}_+}{(r)}
&=&
-\frac{\mathfrak{b}(\sigma)\left(1-\frac{4{GM}}{3c^2r}\right)}{9G^2  r^3\left(1-\frac{3{GM}}{2c^2r}\right)^2\sigma},\\
{{}{\mathcal{U}}_+}{(r)}
&=&
\frac{\mathfrak{b}(\sigma){M}}{12G c^2 r^4\left(1-\frac{3{GM}}{2c^2r}\right)^2\sigma}\ .
\ea
In the vacuum, $p(r)=\uprho(r)=0$ in the $r>{\rm R}$ outer sector.  Therefore, the outer  metric can be written as \cite{Casadio:2012rf,Ovalle:2016pwp}
\begin{equation}
\label{genericext}ds^2\!=\!-e^{\bm{\upnu}^+(r)} dt^2\!+\frac{dr^2}{1\!-\!\frac{2GM}{c^2r}\!+\!\tilde{h}(r)}+r^{2} d\Upomega^2.
\end{equation}
Junction conditions that match the outer to the inner stellar sector, at the stellar configuration surface $r={\rm R}$, imply that  \cite{Casadio:2012rf}
\begin{eqnarray}
 \label{ffgeneric1}
{\bm{\upnu}^\pm({\rm R})}&=&\ln\left(1-\frac{2{GM}}{c^2{\rm R}}\right),
\\
 \label{ffgeneric2}
 M-M_{\scalebox{.6}{Schw}}&=&\frac{{\rm R}}{2G}\left(\tilde{h}({\rm R},\sigma)-\tilde\upkappa({\rm R},\sigma)\right).
\end{eqnarray}
Regarding Eq. (\ref{ffgeneric2}), the ADM mass reads $M=M_{\scalebox{.6}{Schw}}+{\cal O}(\sigma^{-1})$. Owing to the current brane tension lower bound $\sigma \approxeq 2.813\times 10^{-6} \;{\rm GeV}$ \cite{Fernandes-Silva:2019fez}, all terms involving ${\cal O}(\sigma^{-2})$ are disregarded. 
The junction conditions at the star surface $r={\rm R}$ yield  
\be
\label{matching1}
\dot\lfloor{{g}}_{\mu\nu}\,a^\nu\dot\rfloor=0,
\ee
where $a^\alpha$ denotes a radial vector field and  one denotes the matching function by 
\beq
\dot\lfloor\upkappa\dot\rfloor\equiv\lim_{r\to {\rm R}_-}\upkappa(r)-\lim_{r\to {\rm R}_+}\upkappa(r),\eeq
averaging any quantity on the star surface by its values in both the inner and the outer surface neighbourhoods.  

Thus, Eq.~\eqref{matching1} implies $\dot\lfloor T^{{}^\intercal \mu\nu}\,a_\nu\dot\rfloor=0$, yielding \cite{Ovalle:2013vna}
\be
\label{matching3}
\dot{\Big\lfloor}
2\left[\sigma+\uprho(r)\right] p(r)+\uprho^2(r)
+{4G^2}\left[\mathcal{U}(r)+2G^2\mathcal{P}(r)\right]
\dot{\Big\rfloor}=0
\ .
\ee

The Schwarzschild-type coefficient $e^{\bm{\upnu}_{\scalebox{.6}{Schw}}(r)}=e^{-\upxi_{\scalebox{.6}{Schw}}(r)}
=
1-\frac{2{GM}}{c^2r}$ can be reinstated in Eq.~\eqref{def}, implying that  
\be
\label{defS}
\tilde{h}(r)=
-\frac{2(1-\frac{2GM}{c^2r})\mathfrak{b}(\sigma)}{\left(r-\frac{3GM}{2c^2}\right)r}.
\end{equation}
To derive the function $\mathfrak{b}(\sigma)$, firstly  Eq.~\eqref{matching3} must be rewritten as \cite{Casadio:2012rf}, 
\be
\label{sfgeneric}
{\rm R}^2p({\rm R})-{G^2\tilde\upkappa({\rm R})}\left({{\rm R}\bm{\upnu}'({\rm R})}+1\right)
=
-\tilde{h}({\rm R}).
\ee
It means that the function 
$\tilde{h}({\rm R})$ -- evaluated at the star surface -- attains negative values. Therefore, the MGD coordinate singularity, $r_{\scalebox{.5}{MGD}}=2{GM}/c^2$, is spotted near the center of the MGD stellar configuration, when compared to the Schwarzschild coordinate singularity $r_{\scalebox{.6}{Schw}}=2GM_{\scalebox{.6}{Schw}}/c^2$. In fact, recall that $M=M_{\scalebox{.6}{Schw}}+{\cal O}(\sigma^{-1})$. Therefore, Weyl fluid effects on the brane induce a weaker gravitational force, when compared with standard Schwarzschild solutions \cite{Casadio:2012rf,Ovalle:2007bn}.

Eqs. (\ref{defS}, \ref{sfgeneric}) imply that   \cite{Casadio:2012rf}
\begin{eqnarray}
\label{beta}
\mathfrak{b}(\sigma)
&=&\frac{\frac{3\,{GM}}{2}-c^2{\rm R}}{2GM-c^2{\rm R}}
\left[{\rm R}^3p({\rm R})-{\rm R}\left({{\rm R}\bm{\upnu}'({\rm R})}{}+1\right)G{\tilde\upkappa({\rm R})}\right]\nonumber\\
&\equiv& \frac{\mathfrak{d}_0}{\sigma},
\end{eqnarray}
{}{where $\mathfrak{d}_0$ is given by the awkward expression in Eq. (31) in Ref. \cite{Casadio:2013uma}}.

 Ref. \cite{Casadio:2015jva}  derived the corresponding experimental and observational signatures of a bulk Weyl fluid, obtained from the Solar system classical tests, encompassing the perihelion precession of Mercury, the deflection of light by the Sun
, and the radar echo delay. The bound   
$\Big|\frac{\mathfrak{d}_0}{\sigma}\Big|\lesssim 2.8 \times 10^{-11}$ has been obtained in Ref. \cite{Casadio:2015jva}.

The MGD metric can be then written as  \cite{Casadio:2012rf}
\begin{subequations}
\ba
\label{nu}
\!\!\!\!\!\!A(r)
&=&e^{\bm{\upnu}_{\scalebox{.6}{Schw}}(r)}=
1-\frac{2GM}{c^2r}
\ ,
\\
\!\!\!\!\!\!B(r)
&=&
A(r)\left(\frac{{}{\bm{\ell}}}{r-\frac{3GM}{2c^2}}+1\right)
\ ,
\label{mu}
\ea
\end{subequations} 
where 
\begin{equation}
\label{L}
{}{{}{\bm\ell}
=
{\rm R}\left({\rm R}-\frac{3GM}{2{c^2}}\right)\left({\rm R}-\frac{2GM}{c^2}\right)^{\!-1}\!{}{}{\frac{\mathfrak{d}_0}{\sigma}}}.
\end{equation} 
In  the GR, rigid brane  
limit $\sigma\to\infty$, the MGD metric is clearly lead to the Schwarzschild metric, as $M=M_{\scalebox{.6}{Schw}}+{\cal O}(\sigma^{-1})$.  
Hence, using  the classical tests of GR and replacing the bound   
$\Big|\frac{d_0}{\sigma}\Big|\lesssim 2.8 \times 10^{-11}$ into Eq. (\ref{L}), with units
$c=2.998\times 10^{8}~{\rm m/s}$, $M_{\odot}=1.989\times 10^{30}~{\rm kg}$, 
${\rm R}_{\odot}=6.955 \times 10^8~ {\rm m}$,  yields  the fundamental gravitational length in the MGD metric (\ref{mu}) to be 
\beq\label{elll}
|\,\bm{\ell}\,|\lesssim 6.259\times 10^{-4}\,{\rm m}.\eeq 
This universal bound holds for MGD compact stellar configurations of any mass. In fact, varying 
the mass in Eq. (\ref{L}) also makes the star surface radius, ${\rm R}$, and hence the functions in Eq. (\ref{beta}) to be modified, accordingly.

\section{MGD anomalies}
\label{dsfb}
%Here MGD black holes are studied and their trace anomalies are computed, and compared to the standard Schwarzschild results.

\par As stated briefly in the introduction, our goal is to compare the trace anomalies from the field theory side against the one found in the CFT. This was introduced in Ref. \cite{Casadio:2003jc} and consists in defining the coefficient
\begin{equation}\label{gammaCFT}
	\Upupsilon_{\scalebox{.63}{CFT}}=\left|\frac{\left\langle T\right\rangle _{\scalebox{.63}{4D}}-\left\langle T\right\rangle _{\scalebox{.63}{CFT}}}{\left\langle T\right\rangle _{\scalebox{.63}{CFT}}}\right|\ ,
\end{equation}
where\footnote{Here $R_{\mu\nu}$ is the Ricci tensor and $R$ the scalar curvature.}
\begin{align}
		\left\langle T\right\rangle _{\scalebox{.63}{4D}}&=\frac{1}{2880\pi^{2}}\left(\mathring{K}-R^{\mu\nu}R_{\mu\nu}-\Box R\right), \label{T4D} \\
	\left\langle T\right\rangle _{\scalebox{.63}{CFT}}&=\frac{1}{4\ell_{p}^{2}\sigma^{2}}\left(R^{\mu\nu}R_{\mu\nu}-\frac{1}{3}R^{2}\right), \label{TCFT}
\end{align}
 are the trace and holographic Weyl anomalies, both normalized 
respectively to the number of bosons in the boundary and the 
number of stack branes.
As the form of the Weyl anomaly is different for
different types of matter fields, one has to sum over all massless fields of spin less than two. They also arise by momentum-space dimensional regularization. A counterterm that must be added to eliminate divergences in the contributions of the field to the graviton self-energy  \cite{Christensen:1977jc,Birrell:1982ix}. The quantity $\left\langle T\right\rangle _{\scalebox{.63}{4D}}$ is obtained using only field theoretic methods in curved spacetimes \cite{Birrell:1982ix}, and does not vanish in general for curved backgrounds (even for Ricci flat ones), as it depends on the Kretschmann scalar, $\mathring{K}$. The holographic Weyl anomaly appears when one evaluates the effective action of a CFT, via the AdS/CFT procedure. When computing the effective action of the boundary theory on the brane one is forced to choose one amongst the equivalence class of metrics forming the conformal structure of the boundary, therefore explicitly breaking the conformal symmetry to obtain a finite value. This anomaly is perceived as an UV effect, since it is present in the boundary theory, but arises from a divergence whose origin is on the infrared (IR) scale. Such a divergence is present in the bulk \cite{Henningson:1998gx}.
\par The explicit form of this anomaly depends on the dimension of the spacetime where the CFT boundary is placed. For odd dimensions, the anomaly always vanishes, whereas for even dimensionality the expressions get more intricate as the number of dimensions increase \cite{Henningson:1998gx}. We only present the  4D case, which is the case we are interested in, the anomaly reads 
\begin{equation}
	\mathcal{\mathcal{A}}=-\frac{N^{2}}{\pi^{2}}\left(E_{\left[4\right]}+I_{\left[4\right]}\right),
\end{equation}
considering  a stack of $N$ branes, 
where $E_{\left[4\right]}$ is the Euler density and $I_{\left[4\right]}$ is the conformal invariant.  
This result holds in the context of 5D asymptotic AdS backgrounds \cite{Henningson:1998gx} and here we consider the possibility of  emulating this result to the MGD metric.

 In 4D, there is only one conformal invariant, which is the Weyl tensor contracted with itself. The explicit expressions for the invariants are
\begin{align}
	\begin{aligned}\label{inv10}
	E_{\left[4\right]}&=\frac{1}{64}\left(K-4R^{\mu\nu}R_{\mu\nu}+R^{2}\right),\\I_{\left[4\right]}&=-\frac{1}{64}\left(K-2R^{\mu\nu}R_{\mu\nu}+\frac{1}{3}R^{2}\right).
	\end{aligned}
\end{align}
It is clear from Eq. (\ref{inv10}) that $E_{\left[4\right]}+I_{\left[4\right]}=-\frac{1}{32}\left(R^{\mu\nu}R_{\mu\nu}-\frac{1}{3}R^{2}\right)$. Eq. \eqref{TCFT} is then obtained by using the AdS/CFT dictionary on braneworlds, that relates $N$ degrees of freedom to the Planck length and brane tension \cite{Soda:2010si,Casadio:2003jc}.

%$\left\langle T\right\rangle _{\scalebox{.63}{CFT}}$, on the other hand, has its source on bulk perturbations - which is the source of the energy-momentum tensor on the boundary CFT [ref needed here?].

{}{The developments in the beginning of Sect. \ref{MGD} can be now equivalently implemented in the context of AdS/CFT. Firstly, the brane Einstein 
equations can be expressed as \cite{Randall:1999vf,Shiromizu:2001jm,deHaro:2000wj,Gubser:1999vj,Henningson:1998gx}:
\begin{eqnarray}
\!\!\!\!\!\!\!\!\!\! {\rm G}_{\mu\nu}=\frac{8\pi G}{c^4}T_{\mu\nu}+\frac{4}{{a|-g|^{1/2}}}
\frac{\delta}{\delta g_{\mu\nu}}\left(\Upgamma_{\scalebox{.6}{CFT}}+\frac{1}{32}S_{\circ}\right),
\label{19}
\end{eqnarray} for $a=4K^{-1}$, where $K=K^\mu_{\;\mu}$ is the extrinsic curvature trace. In addition, the quantity $\Upgamma_{\scalebox{.6}{CFT}}$ carries the CFT action on the
boundary. Its trace anomaly reads \cite{deHaro:2000wj,Henningson:1998gx}:
\begin{eqnarray}
g_{\mu\nu}\frac{\updelta \Upgamma_{\scalebox{.6}{CFT}}}{\updelta g^{\mu\nu}}=
\frac{a^3}{16}
{|-g|^{1/2}}\left(
R^{\mu\nu}
R_{\mu\nu}-{\frac{1}{3}}
R^2  \right). \label{eq:anomaly}
\end{eqnarray}
The term 
$S_\circ$
 has $R^2$ counterterms that yield a finite action, whereas the trace of the term $\updelta S_\circ/\updelta g_{\mu\nu}$ equals zero, for  
\begin{eqnarray}
\frac{\delta S_\circ}{\delta g^{\mu\nu}}
&\! = &\! \frac{a^3}{2}
\left[\Box R_{\mu\nu}\!-\frac{1}{3}\nabla_\mu \nabla_\nu R-2R^{\upalpha\upbeta}R_{\mu\upalpha\nu\upbeta}\!+\!\frac{2}{3}RR_{\mu\nu}\right.\nonumber\\
&&\left.\qquad -\frac{1}{2}g_{\mu\nu}\left(\frac13(\Box R\!+\!R^2) \!-\!\frac{1}{4}R^{\upalpha\upbeta}R_{\upalpha\upbeta}
\right)
\right],
\end{eqnarray}
where $\upalpha,\upbeta=0,1,2,3$. 
Now, taking the trace of all terms in Eq. (\ref{19}) yields $
R=-\frac{8\pi G}{c^4} T+\frac{a^2}{4}
\Bigl({\textstyle \frac{1} {3}}
R^2-
R^{\mu\nu}
R_{\mu\nu}  \Bigr)$. Therefore, up to linear order,  the CFT stress-energy-momentum tensor emulates the brane Weyl tensor part, that can be read off as \cite{Randall:1999vf,Gubser:1999vj,deHaro:2000wj}
\begin{eqnarray}
E_{\mu\nu} = - \frac{K}{{|-g|^{1/2}}}
\frac{\updelta \Upgamma_{\scalebox{.63}{CFT}}}{\updelta g^{\mu\nu}}.
\end{eqnarray}
Hence, it appropriately locates the on-brane Weyl tensor (\ref{A4}), and its consequences to the Einstein effective brane equations (\ref{5d4d}), in the AdS/CFT setup.}

\par The coefficient \eqref{gammaCFT} measures the reliability of results obtained through the AdS/CFT correspondence when a given spacetime metric is under investigation in the following sense: by measuring the trace anomalies, one can check how the back-reaction of bulk perturbations affect results on the brane or vice-versa, that is if the presence of the brane has any effect on the bulk geometry. The value of $\Upupsilon_{\scalebox{.63}{CFT}}$ ranges from $0$ to infinity, and the predictions from AdS/CFT become less reliable the higher the value of the coefficient is \cite{Casadio:2003jc}.

\par If one considers the Schwarzschild spacetime \cite{Casadio:2003jc}, setting $\bm{\ell}=0$ in Eq. \eqref{mu}, the immediate results $\left\langle T\right\rangle _{\scalebox{.63}{4D}}\propto K$ and $\left\langle T\right\rangle _{\scalebox{.63}{CFT}}=0$ are found. The second identity leads to the  conclusion that $\Upupsilon_{\scalebox{.63}{CFT}}\to\infty$, and therefore results from the AdS/CFT correspondence are, at best, questionable.

\par For the MGD metric \eqref{abr} with functions $A(r)$ and $B(r)$ given by Eqs. \eqref{nu} and \eqref{mu}, respectively, one can thoroughly compute the anomalies, as well as the $\Upupsilon_{\scalebox{.63}{CFT}}$ coefficient, as 
\begin{widetext}
	\beq\label{T4DMGD}
	\left\langle T\right\rangle^{\scalebox{.63}{MGD}} _{\scalebox{.63}{4D}}&=&\frac{1}{360\pi^{2}r^{6}\left(3cGM-2c^{2}r\right)^{2}}\left[486G^{6}M^{6}+9c^{12}{\bm{\ell}}^{2}r^{4}-648c^{2}G^{5}M^{5}\left({\bm{\ell}}+2r\right)-16c^{10}G{\bm{\ell}}Mr^{3}\left(5{\bm{\ell}}+3r\right)\right.\nonumber\\&&\left.+216c^{4}G^{4}\left({\bm{\ell}}+r\right)\left({\bm{\ell}}+6r\right)-36c^{6}G^{3}M^{3}r\left(12{\bm{\ell}}^{2}+35{\bm{\ell}}r+16r^{2}\right)+6c^{8}G^{2}M^{2}r^{2}\left(49{\bm{\ell}}^{2}+72r{\bm{\ell}}+16r^{2}\right)\right]\\
	\left\langle T\right\rangle _{\scalebox{.63}{CFT}}^{\scalebox{.63}{MGD}}&=&\frac{{\bm{\ell}}^{2}}{\sigma^{2}\ell_{p}^{2}}\frac{2c^{2}\left(6G^{2}M^{2}-8c^{2}GMr+3c^{4}r^{2}\right)}{r^{4}\left(3GM-2c^{2}r\right)^{4}}\\\label{gammaMGD}
	\Upupsilon_{\scalebox{.63}{CFT}}^{\scalebox{.63}{MGD}}&=&\left|1-\frac{\sigma^{2}\ell_{p}^{2}}{720\pi^{2}c^{8}{\bm{\ell}}^{2}r^{2}\left(3c^{4}r^{2}-8c^{2}GMr+6G^{2}M^{2}\right)}\left[9c^{12}{\bm{\ell}}^{2}r^{4}-16c^{10}G{\bm{\ell}}Mr^{3}(5{\bm{\ell}}+3r)\right.\right.\nonumber\\&&+6c^{8}G^{2}M^{2}r^{2}\left(49{\bm{\ell}}^{2}+72{\bm{\ell}}r+16r^{2}\right)-36c^{6}G^{3}M^{3}r\left(12{\bm{\ell}}^{2}+35{\bm{\ell}}r+16r^{2}\right)\nonumber\\&&\left.\left.+216c^{4}G^{4}M^{4}\left({\bm{\ell}}^{2}+7{\bm{\ell}}r+6r^{2}\right)-648c^{2}G^{5}M^{5}({\bm{\ell}}+2r)+486G^{6}M^{6}\right]\right|,
\eeq
\end{widetext}
where $\ell_p=\sqrt{\hbar G/c^3}$ denotes the Planck length. In the large-$r$ limit, Eq. \eqref{T4DMGD} reads
\begin{equation}
	\left\langle T\right\rangle _{\scalebox{.63}{4D}}^{\scalebox{.63}{MGD}}=\left(1-\frac{Gc^{2}M}{120}{\bm{\ell}}+\frac{c^{4}}{640}{\bm{\ell}}^{2}\right)\left\langle T\right\rangle _{\scalebox{.63}{4D}}^{\scalebox{.63}{Schwarzschild}}
\end{equation}
that is, it constitutes corrections to the trace anomaly for the Schwarzschild black hole solution. This motivates to analyze Eq. \eqref{gammaMGD} on the same regime, where we obtain
\begin{equation}\label{gammar0}
	\!\!\!\!\!\Upupsilon_{\scalebox{.63}{CFT}}^{{\scalebox{.63}{MGD}}({r\to\infty})}\!=\!\left|\frac{\sigma^{2}\ell_p^{2}\left(3c^{4}l^{2}\!-\!16c^{2}G{\bm{\ell}}M\!+\!32G^{2}M^{2}\right)}{720\pi^{2}c^{4}{\bm{\ell}}^{2}}\!-\!1\right|
\end{equation}
Eq. \eqref{gammar0} allows us to draw some conclusions about the reliability of bulk/boundary correspondence, as long as one bears in mind that such conclusions apply for $r\gg1$.
\par We proceed to extract numerical values for $\Upupsilon_{\scalebox{.63}{CFT}}$ for two known values of the ${\bm{\ell}}$ parameter, that appears 
in the MGD metric radial component (\ref{mu}),  respectively  for two regimes of the ADM mass $M$ \cite{Casadio:2015jva,Cavalcanti:2016mbe}. In the following calculations, we will use SI units for all quantities\footnote{$\ell_{p}=1.616\times 10^{-35}$m, $G=6.674\times 10^{-11}$m$^{3}$kg$^{-1}$s$^{-2}$.}. Ref. \cite{Casadio:2013uma} has shown that {${\bm{\ell}}\sigma=-0.0042572$}.
%, where $R$ is the radius of the stellar distribution or the horizon event radius, which is then simply $R=2M$. 
This is particularly interesting because we can then eliminate the brane tension in Eq. \eqref{gammar0} and write it only in terms of known quantities:
{{}{\begin{equation}\label{gammaNsigma}
	\!\!\!\!\Upupsilon_{\scalebox{.53}{CFT}}^{{r\to\infty}}\!=\!\left|\frac{ 10^{-6}\ell_p^{2}\!\left(3c^{4}{\bm{\ell}}^{2}\!-\!16c^{2}G{\bm{\ell}}M\!+\!32G^{2}M^{2}\right)}{4\pi^{2}c^{4}{\bm{\ell}}^{4}}\!-\!1\right|.
\end{equation}}}
\par The first set of data to be considered is the already \blt{upper bound}  (\ref{elll}) \cite{Casadio:2015jva}, the other relevant values are given above that equation. Besides, in Ref. \cite{Cavalcanti:2016mbe} the MGD solution was applied to modelling gravitational lensing effects, where the Sagittarius A${}^*$ black hole of mass $M=(4.154\pm0.014)\times 10^{6}M_{\odot}$ was considered, therefore ${\rm R}=2M$ is the event horizon radius. For this case, the observational  value of ${\bm{\ell}}=0.06373\; {\rm m}$ was  obtained  \cite{Cavalcanti:2016mbe}. For both sets, $|{\bm{\ell}}|\ll 1$ leading  Eq. \eqref{gammaNsigma} to imply that \beq
\Upupsilon_{\scalebox{.63}{CFT}}^{\scalebox{.63}{MGD}}\approx 1.\eeq

 When ${\bm{\ell}}\to 0$, all the results regarding the Schwarzschild solution are recovered, however the current range of the brane tension, $\sigma \gtrapprox  2.813\times10^{-6} \;{\rm GeV}$ \cite{Fernandes-Silva:2019fez}, together with  the equality {}{${\bm{\ell}}\sigma=-0.0042572$}, makes $\ell$ not to attain a null value. Therefore,
 it places the MGD, in the AdS/CFT formulation on the brane, into a trustworthy position to be a realistic model to describe, in this context, stellar distributions that are compatible with AdS/CFT. 
\blt{Although the most precise lower bound that the brane tension attains $\sigma \gtrapprox  2.813\times10^{-6} \;{\rm GeV}$ \cite{Fernandes-Silva:2019fez} can be used, it is worth to mention the physically feasible and realistic  value $\sigma\sim 7.3 \times 10^{19}
{\rm kg.m^2/s^2} = 0.3 \;{\rm GeV}$ was reported in Refs. \cite{Germani:2001du,Boehmer:2008zh}. This value was also found in the membrane paradigm describing neutron stars, which although  stronger than the big-bang nucleosynthesis constraint probed by the Cosmic Background Explorer \cite{Maartens:1999hf}, is still weaker than the bound obtained by Newton's law at
submillimetric scale constraints. Therefore this value can be realistically employed for our purposes here}.

\par \blt{Using strong gravitational lensing effects of anisotropic stellar configurations, in particular the catalogued PSR B0943+10,   4U 1820.30, 4U 1728.34, PSR J0348+0432, RX J185635.3754, PSR J0348+0432, PSR 0943+10, SAX J1808.4-3658, Her X-1, Cen X-3, XTE J1739-285,  RX J1856, XTE J1739-285, Sagittarius A* \cite{Cavalcanti:2016mbe,Maurya:2019noq,Tello-Ortiz:2019gcl,Torres:2019mee,Deb:2018ccw}, including also the possibility that our sun may be described by the MGD as posed in Ref.  \cite{Casadio:2015jva}, we can then engender the more general relation 
\beq
	{\bm{\ell}}(M)&=&1.95243\times 10^{-75} M^2 - 7.84876\times10^{-39} M \nonumber\\&&\qquad\qquad\qquad\qquad\qquad+ 0.06373, 
\eeq with mass $M$ units in kg. Applying this expression to \eqref{gammaMGD} and taking the limit $r\to 2M$ yields }
\begin{widetext}
\beq \label{gammar2M}
\!\!\!\!\!\!\!\!\!\!\!\!\blt{\lim_{r\to2M}\Upsilon_{\scalebox{.53}{CFT}}=\left|\frac{-3.812\times10^{-150}M^{4}+3.065\times10^{-113}M^3-3.105\times10^{-76}M^{2}+1.001\times 10^{-39}M-4.001\times10^{-3}}{\left(6.373\times10^{-2}-7.849\times10^{-39}M+1.952\times10^{-75}M^{2}\right)^{2}}\right|.}
\eeq 
\end{widetext}
\blt{One can check that $\lim_{r\to2M}\Upsilon_{\scalebox{.53}{CFT}}\approxeq1$ in Eq. \eqref{gammar2M}, for all mass values in the range $0\leq M\lesssim 10^{70}$ kg. Some studies have suggested that the maximum mass that a black hole can reach is $\sim10^{12} M_\odot \approx 10^{42}$ kg \cite{1400,1401}, and that the mass of the entire observable universe is often quoted as $\sim10^{53}$ kg, regarding the ordinary matter that includes the interstellar and the intergalactic media, up to $\sim10^{55}$ kg if one includes  dark matter and dark energy as well. Therefore, we are safe to assert that Eq. \eqref{gammar2M} will be effectively equal to 1, for all values of stellar configuration masses in the range from zero to 17 orders of magnitude more massive than the observable universe. According to Ref. \cite{Casadio:2003jc}, it provides a good agreement for the AdS/CFT correspondence in the IR regime.}

 \par \blt{To summarize, one can safely assume that $\Upupsilon_{\scalebox{.53}{CFT}}$ is effectively unity in both the infinity and the near the event horizon limits, unless one considers either exceptionally high masses that are not physically attainable -- since  one should consider stellar objects 17 orders of magnitude more massive than the entire observable universe -- or models where the brane tension is very high, approaching a rigid brane. The latter case goes towards the GR limit and such behavior is expected.}

\section{Conclusions}
\label{4}
{}{The MGD, usually employed to obtain static, strongly gravitating   spherically symmetric and compact stellar distributions, was here 
explored with the tools of trace and Weyl  anomalies.
Contrary to the Schwarzschild solution, for which the implausible result $\Upupsilon_{\scalebox{.63}{CFT}}\to\infty$ in Eq. (\ref{gammaCFT})  yields the AdS/CFT correspondence to be difficult to be implemented in this context, the MGD is a reliable 
attempt to describe realistic models, in the AdS/CFT setup. 
The parameter in Eq. \eqref{gammaCFT} quantifies how safe  AdS/CFT is when bulk/brane back-reaction effects are taken into account. Since the value of the $\Upupsilon_{\scalebox{.63}{CFT}}$ coefficient, for the MGD case, was shown to be near unity, it means that the MGD solutions may occupy a privileged place and can play a prominent role on emulating AdS/CFT on braneworld scenarios.}

{}{Similarly to AdS/QCD models, where the extra dimension is interpreted as an energy scale in QCD, in the setup here established,  phenomena  regarding CFT coupled to gravity can be exclusively interpreted from the point of view of the brane. Bulk gravitons that propagate in the bulk correspond to 4D gauge bosons on the boundary. The difference between these two countenances, in the AdS/CFT setup, cannot be identifiable, from any phenomenological  point of view. }

\subsection*{Acknowledgements}
This study was financed in part by the Coordenação de Aperfeiçoamento de Pessoal de Nível Superior – Brasil (CAPES) – Finance Code 001.
RdR~is grateful to FAPESP (Grant No.  2017/18897-8 and No. 2021/01089-1) and the National Council for Scientific and Technological Development  -- CNPq (Grants No. 303390/2019-0 and No. 406134/2018-9), for partial financial support.

\bibliography{bib_MGDanomaly}
\end{document}